\documentclass[prl,twocolumn,showpacs,superscriptaddress]{revtex4-1}

\usepackage{graphicx,amssymb,amsmath,color,bm}
\begin{document}
\definecolor{darkgreen}{rgb}{0,0.5,0}

\newcommand{\jav}[1]{\textcolor{red}{#1}}

\title{Topological and trivial magnetic oscillations in nodal loop semimetals}
\author{L\'aszl\'o Oroszl\'any}
\affiliation{Department of Physics of Complex Systems, E\"otv\"os University,
 H-1117 Budapest, Hungary}
\author{Bal\' azs D\' ora}
\affiliation{Department of Theoretical Physics and MTA-BME Lend\"{u}let 
Spintronics
Research Group (PROSPIN), Budapest University of Technology and Economics, 
1521 Budapest, Hungary}
\author{J\'ozsef Cserti}
\affiliation{Department of Physics of Complex Systems, E\"otv\"os University, 
H-1117 Budapest, Hungary}
\author{Alberto Cortijo}
\affiliation{Materials Science Factory, Instituto de Ciencia de Materiales de Madrid, 
CSIC, Cantoblanco; 28049 Madrid, Spain.}
\date{\today}

\begin{abstract}
Nodal loop semimetals are close descendants of Weyl semimetals and possess 
a topologically dressed band structure. We argue by combining the conventional 
theory of magnetic oscillation with topological arguments that nodal loop 
semimetals host coexisting topological and trivial magnetic oscillations. 
These originate from mapping the topological properties of the extremal
Fermi surface cross sections onto the physics of two dimensional semi Dirac
systems, stemming from merging two massless Dirac cones. By tuning the 
chemical potential and the direction of magnetic field, a sharp transition 
is identified from purely trivial oscillations, arising  from the Landau 
levels of a normal two dimensional (2D) electron gas, to a phase where 
oscillations of topological and trivial origin coexist, originating from 2D 
massless Dirac and semi Dirac points, respectively. These could in 
principle be directly identified in current experiments.
\end{abstract}

\maketitle

\paragraph{Introduction.}
Topological nodal semimetals are three dimensional semimetallic systems where 
the valence and conduction bands closest to the Fermi level cross each other 
in momentum space. In the case of Weyl/Dirac semimetals, the crossing consists 
of a discrete set of points, while in the case of nodal loop semimetals (NLSM) 
the crossing takes the form of a closed loop\cite{BurkovBalents11}. While both 
families of semimetals are topologically non-trivial, the nature of the 
topological structure is quite different. In the case of a NLSM, the 1D character
of the line of singularities (that ultimately comes from the discrete symmetries
of the system\cite{FCK15}) determines the topological invariant similar to 1D 
topological insulators\cite{Asbth2016}: Any closed path in momentum space along 
which the Hamiltonian of the system is gapped can be threaded by the loop an odd 
or even (including zero) number of times. In the former case, this closed path 
acquires a Berry phase of $\pi$ while in the latter this Berry phase is zero, 
defining in this way a $Z_{2}$ invariant\cite{FCK15,CHX16}.
A fundamentally intriguing question is whether it is possible to observe this 
$Z_2$ topological character in experimentally accessible properties in NLSMs. 
In the case of Weyl semimetals there is a 
non-trivial Berry curvature (not appearing in NLSM) that directly modifies 
transport and optical properties\cite{AMV17}. For the case of NLSMs, it has been 
suggested that this topological structure should manifest in quantum oscillations
\cite{RK15,LB17,CHX16} or through the presence of surface states\cite{RZS17,GM17}.

The recent theoretical effort to identify NLSM candidates\cite{MUG15,HLV16,XYF17,
QYP17,HOM17,DTW17,lim1}, was accompanied by intense experimental progress, mostly
focused on ARPES \cite{KWK15,YWF15,BCS16,NBH16,SSJ17,WME17}{ yielding surface properties} and magnetotransport
experiments\cite{HTL16,ESD17,ZGZ17,PDS17,MCK17,hu2017}{ sensitive to bulk characteristics}, particularly in the family of 
ZrSi-chalchogenides. 
{Although recent experiments show the presence of a non trivial Berry phase\cite{ZGZ17,PDS17,MCK17,hu2017}, to this day it is unclear how the crossover between trivial and topological oscillations is manifested in magneto oscillation spectra\cite{ASG16}.}
It is thus crucial to develop a theoretical approach to expose the fingerprints of the topological nature of these novel materials.

\begin{figure}[h!]
\includegraphics[width=0.9\linewidth]{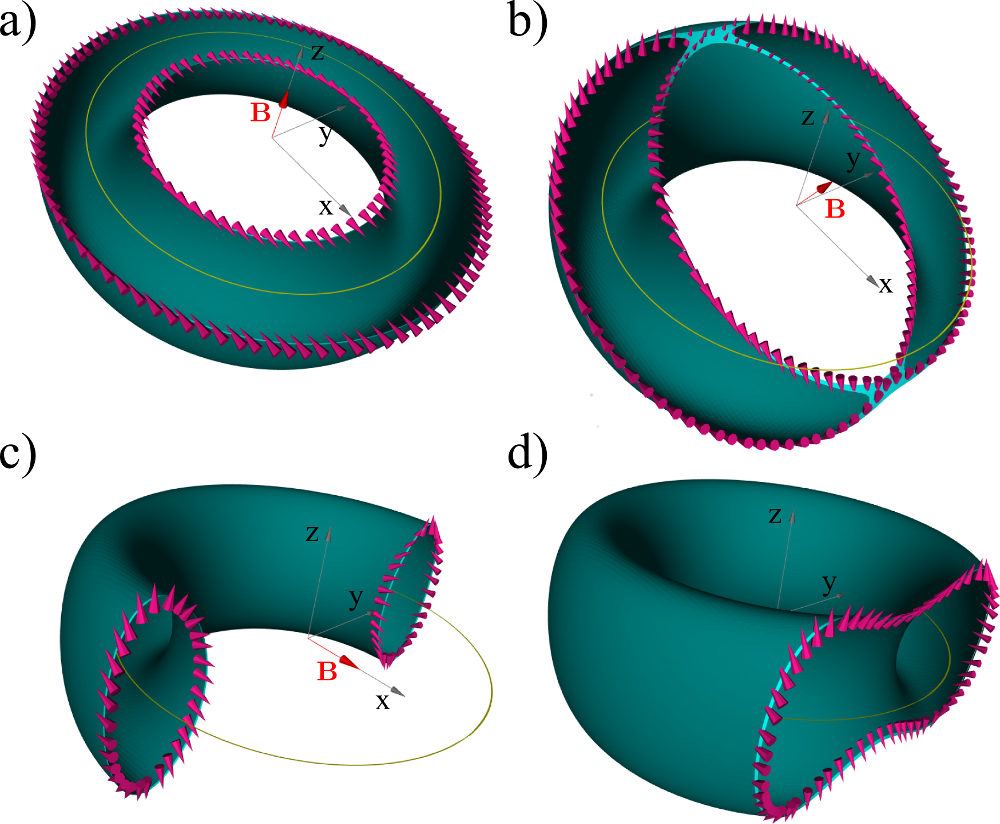}
\caption{The evolution of the cross sections by planes perpendicular to magnetic
field of the torus Fermi surface is shown: a) magnetic field in $z$ direction, 
the winding of  the vector $\bf d(p)$ (magenta cones) reveals topologically 
trivial character.
Panel b) shows the case of critical angle, when the Lifshitz transition occurs. 
The c) and d) panels depict the topological and trivial extremal cross sections 
for magnetic field perpendicular to $z$ axis, the winding of the vector 
$\bf d(p)$ signal $\pm\pi$  and 0 Berry phases, respectively.
}
\label{Bcut}
\end{figure}

In this work we present a comprehensive theoretical description of magnetic 
oscillations present in these novel systems. Based on simple topological 
arguments backed by a semiclassical analysis we construct the phase diagram 
for finite chemical potential and arbitrary field orientation and contrast 
it to numerical calculations. 

\paragraph{Topological content of magnetic oscillations.} 
The effects of an external magnetic field can be qualitatively understood by 
recalling that the electronic motion is confined in the plane perpendicular to 
the magnetic field.  Upon fixing the chemical potential of the electron, it 
follows the trajectory set by the cross section of the
constant energy contour and the plane perpendicular to the magnetic field.
By sweeping the magnetic field, the extremal orbits among these trajectories 
determine the characteristic frequency of magnetic oscillations\cite{abrikosov,
ashcroft}, simply because their contribution dominates over the other orbits.
This gives rise to the celebrated de Haas-van Alphen effect and the 
Shubnikov-de Haas oscillations, which are well documented for normal metals, 
being both quantities intimately related to the evolution of the density of 
states (DOS) when the magnetic field is varied. The quantization of the 
cyclotron orbits is expressed in terms of the Onsager quantization condition
\begin{equation}
\frac{\mathcal{A}[E]\hbar}{eB}=2\pi(n+\gamma),\label{quantizedarea}
\end{equation}
where $\mathcal{A}[E]$ is the area enclosed by the cyclotron orbit in 
momentum space for energy $E$, $n$ is an integer, $B$ is the applied 
magnetic field,
and $\pi(1-2\gamma)$ is the Berry phase accumulated by the cyclotron 
orbit\cite{MS99,glazman}. Eq. \eqref{quantizedarea} implicitly defines the 
Landau level spectrum\cite{abrikosov}.
As mentioned in the introduction, the presence of non trivial Berry phases are 
expected for extremal orbits threading the nodal loop. Thus the $Z_{2}$ 
invariant assigned to the cyclotron orbits can be identified with 
$\gamma$\cite{KWK15}. The presence of cyclotron orbits with non-trivial 
Berry phases has also been linked to the appearance of a almost-flat zero 
energy Landau level\cite{RK15}, similarly to the case of graphene\cite{NGM05,ZTS05}.

\paragraph{Band structure of nodal loop semimetals.} 
We consider the low energy Hamiltonian of a NLSM as
\begin{gather}
H=\left( \Delta-\frac{p_\perp^2}{2m}\right)\sigma_x+
vp_z\sigma_z=\bf d(p)\cdot\bm\sigma,
\label{hamnodalloop}
\end{gather}
where the $\sigma$'s are Pauli matrices, $p_\perp^2=p_x^2+p_y^2$, $m>0$ is 
an effective mass, $\Delta$ is an energy scale, and $v$ is the Fermi velocity in 
the $z$ direction. Note that the orientation pattern of $\bf d(p)$ encodes 
the topological properties of the system. The Hamiltonian can readily be 
diagonalized to yield the spectrum as $E_\pm({\bf p})=\pm\sqrt{v^2p_z^2+
(\Delta-\frac{p_\perp^2}{2m})^2}$.  The absence of $\sigma_y$ signifies the chiral symmetry of the considered model. The presence of this symmetry is required to stabilize the nodal loop in Eq. \eqref{hamnodalloop}.
This also guarantees  the quantized Berry phases.
For $\Delta>0$, the Fermi surface consists of a circle in the $x-y$ plane with 
radius $\sqrt{2\Delta m}$. In the presence of finite chemical potential, 
$\mu>0$, the Fermi surface is determined from the $\mu=\epsilon_+({\bf p})$ 
relation, and the Fermi surface evolves from a circle at $\mu=0$ to a torus 
like surface for finite $\mu$, whose tube radius is set by the chemical 
potential. For the present model in Eq. \eqref{hamnodalloop}, two qualitatively 
different characteristic cases arise for the quantum oscillations by considering
a magnetic field parallel and perpendicular to the $z$ axis.

\paragraph{Extremal Fermi surface cross sections.} 
First we discuss the magnetic field in the $z$ direction, thus restricting 
electrons to the $x-y$ plane. The extremal cross section of the Fermi surface 
is depicted in Fig \ref{Bcut} a). In this case the effective Hamiltonian, 
Eq. \eqref{hamnodalloop} is dominated by its first term, $\bf d(p)$ points 
always in the $x$ direction hence the spinor structure is prevented from 
acquiring any winding. The emerging magnetic oscillations are thus doomed 
to be trivial with Berry phase 0 and  $\gamma=1/2$. 

\begin{figure}
\includegraphics[width=0.47\linewidth]{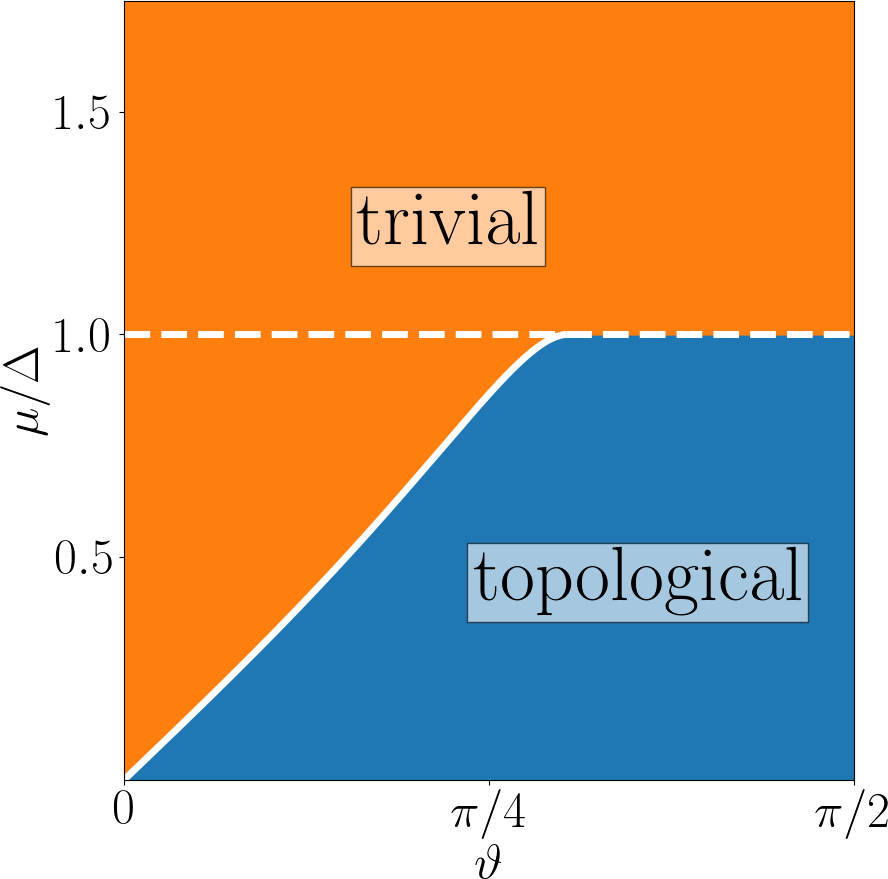}
\includegraphics[width=0.47\linewidth]{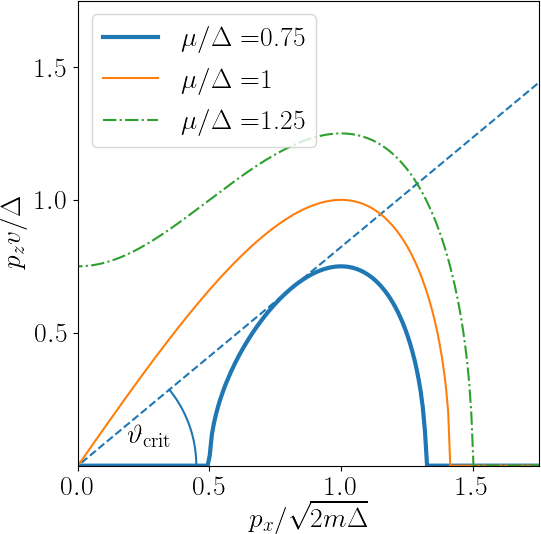}
\caption{The phase diagram of Eq. \eqref{hamnodalloop} is shown as a function 
of energy (chemical potential) and magnetic field angle (left panel) for 
$\Delta=2v^2m$. At a fixed small energy, the oscillations are non-topological 
for small tilt angles and become a mixture of topological and trivial 
oscillations upon tilting the magnetic field further from the $z$ axis.
The right panel,  visualizing a quarter of a crossection of the Fermi 
surface, depicts the geometric interpretation of the critical tilt 
angle in the $p_y=0$ plane, $\vartheta_{crit}$ growing  with the chemical 
potential. When the Fermi surface becomes a horn torus at $\mu=\Delta$,
only topologically trivial cross sections exits.
}
\label{phasediag}
\end{figure}

Turning the magnetic field perpendicular to the $z$ axis, the effective motion 
of electrons is confined into vertical cuts through the torus: depending on 
the position of the cutting plane, cross sections can either be a single loop 
or two disconnected closed rings. The resulting physics is dictated by a semi 
Dirac point in 2D\cite{montambaux,montambaux2009} with an effective Hamiltonian 
in the $y-z$ plane as 
\begin{gather}
H_{eff}=\left(\tilde \Delta-\frac{p_y^2}{2m}\right)\sigma_x+vp_z\sigma_z,
\end{gather}
where $\tilde \Delta=\Delta-\frac{p_x^2}{2m}$ and the magnetic field points in 
the $x$ direction. For $\tilde \Delta=0$, a semi Dirac point is realized with 
a combination of quadratic and linear dispersions (in the $y$ and $z$ 
directions), respectively.
When $\tilde \Delta<0$, the spectrum is gapped, the dispersion above the gap is 
reminiscent to that of a 2D anisotropic mormal electron gas, thus
topologically trivial.
Finally, the most important situation from a topological point of view arises 
for $\tilde \Delta>0$, when the cross section by a plane perpendicular 
to the magnetic field hosts two linearly dispersing Dirac cones, 
carrying a Berry phase of $\pm\pi$.

For a given chemical potential, the extremal cross sections occur in two 
different locations of the cutting plane: there is a single, connected, 
topologically trivial, maximal Fermi surface and there are two, disconnected, 
topological (due to the $\pm\pi$ Berry phases), minimal Fermi surfaces, 
visualized in Fig. \ref{Bcut} d) and c) respectively.
These Fermi surface sections determine the magnetic oscillations, which will 
be a mixture of topological and non-topological frequencies, stemming from 
the aforementioned disconnected and connected extremal Fermi surfaces, 
respectively.
Their topological content is revealed by following the winding of vector 
$\bf d(p)$ in Eq. \eqref{hamnodalloop} upon going around a closed cross section.
In the case of the connected, maximal cross section, we can unwind $\bf d(p)$ 
to point in the same direction for all $\bf p$ around the cross section 
by plane, thus representing a topologically trivial surface.
On the other hand, for the two disconnected, minimal area Fermi surfaces
cross sections, the winding of $\bf d(p)$ is identical to that of 
graphene\cite{hasankane,castro07}, going around clockwise and anticlockwise 
in the two disconnected cross sections, giving rise to Berry phases $\pm\pi$.

\begin{figure}
\includegraphics[width=\linewidth]{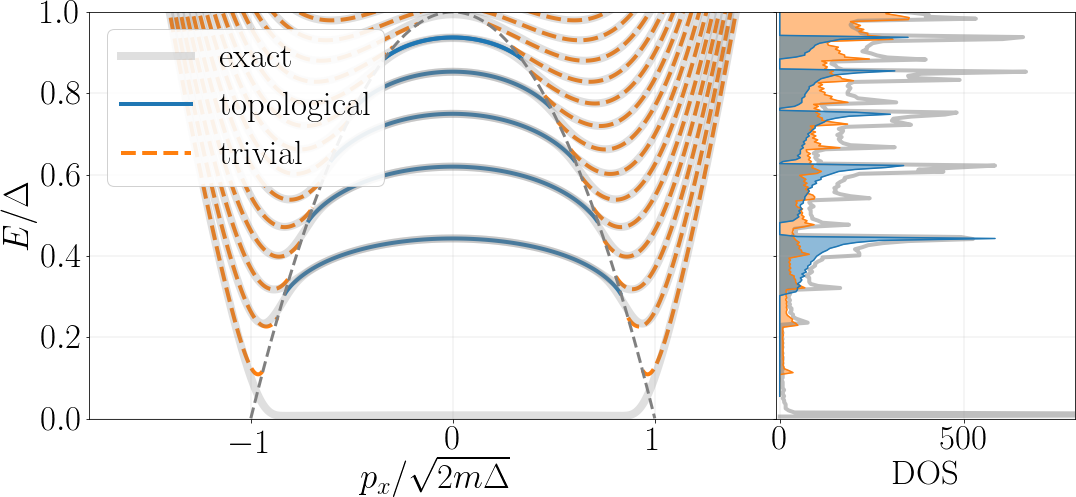}
\caption{Spectrum of the in plane magnetic configuration obtained by exact 
diagonalization and by the semiclassical approach for $\sqrt{2m}\Delta^{3/2}/veB
=20$. 
The dashed gray lines signal the boundary between topological and trivial 
part of the spectrum. The corresponding DOS from numerics as well as the 
separate topological and trivial contributions from semiclassics is plotted, 
and is dominated by the topological contribution at small energies.}
\label{inplanespectrum}
\end{figure}

For a general magnetic field tilted by an angle $\vartheta$ from the $z$ axis, 
there is a sharp transition between only non-topological and a combination of 
topological and non-topological magnetic oscillations.
For small $\vartheta$, the extremal Fermi surfaces are still topologically 
trivial: the extremal Fermi surfaces corresponding to  $\vartheta=0$ are 
adiabatically connected to the case of small finite tilting angle,
therefore their topologically trivial nature remains unchanged.
At a critical angle
\begin{gather}
\tan(\vartheta_{\mathrm{crit}})=(v\sqrt{m})^{-1}\sqrt{\Delta-
\sqrt{\Delta^2-\mu^2}},
\label{criticaltheta}
\end{gather}
a Lifshitz transition occurs and for larger angles, the extremal Fermi surfaces
are  qualitatively similar to the 2D semi Dirac case.
These are adiabatically connected to the $\vartheta=\pi/2$ case and contain 
topological (with Berry phase $\pm\pi$) and non-topological (with Berry phase 
$0$) features.
Finally, for large chemical potential, another transition occurs and the 
\emph{ring} torus shaped Fermi surfaces in Fig. \ref{Bcut}  change into 
\emph{spindle} torus, in which case even the previously disconnected Fermi 
surface loops touch and become topologically trivial. All these features are 
summarized in Fig. \ref{phasediag}. Note that the quantization of the Berry phase to $\pi$ and $0$ is guaranteed by the same symmetry that protects the zero energy nodal line. 

\begin{figure*}[t!]
\includegraphics[width=\linewidth]{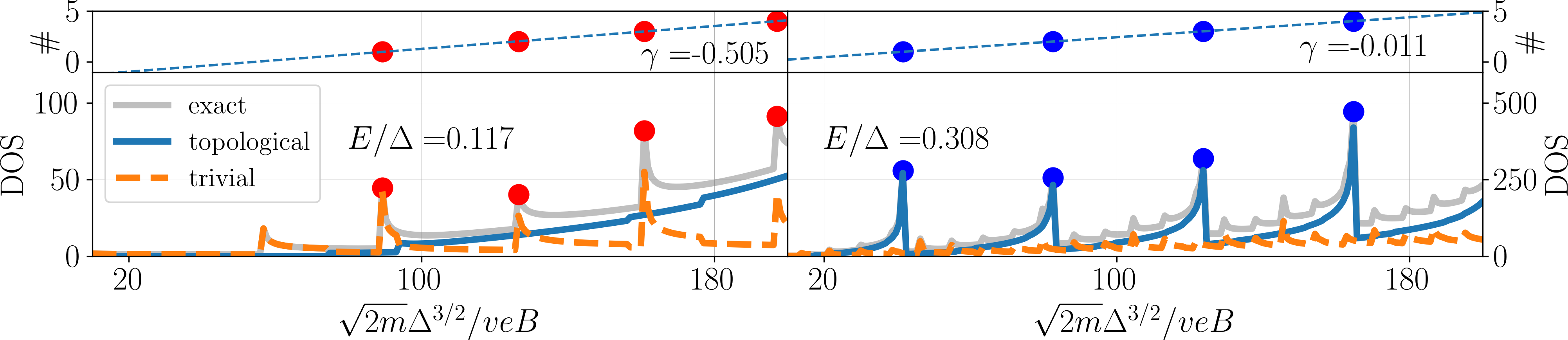}
\caption{Magnetic oscillations obtained by the numerical solution of 
Eq. \eqref{hamnodalloop} (exact DOS), as well as from the Onsager approach 
(topological and trivial DOS) as the function of the inverse magnetic field
parallel to the $x$ axis with numerically extracted $\gamma$. 
For small energies the oscillations of this magnetic interval are dominated by 
trivial oscillations (left panel) while either increasing energy (right panel) 
or decreasing the magnetic field the oscillations will be dominated by 
oscillations of topological origin. Note that by working at a fixed particle 
number as opposed to a fixed energy, $\gamma$ is expected\cite{murakawa} to be 
shifted by an additional $\pm 1/8$.}
\label{inplaneosc}
\end{figure*}

\paragraph{Semiclassical analysis.} 
Having determined the general structure in magnetic field oscillations from the 
extremal Fermi surface cuts, we turn to a more quantitative analysis by inserting
a finite magnetic field into Eq. \eqref{hamnodalloop}. The orbital effects of an
external magnetic field in the $x-z$ plane are captured by the vector potential 
$\mathbf{A}=B\left (0,\cos(\vartheta)x-\sin(\vartheta)z ,0\right )$ 
where $\vartheta$ is measured from the $z$ axis, using the Peierls substitution,
${\bf p\rightarrow p}-e{\bf A}$ in Eq. \eqref{hamnodalloop}~\footnote{For all 
magnetic field orientations, $p_y$ only constitutes a trivial shift of the 
cyclotron orbits but the spectrum will not depend on it.}.
For a magnetic field in the $z$ direction, $\vartheta=0$, the spectrum reads 
as\cite{LB17} $E_{\pm,n,p_z}=\pm\sqrt{( \Delta-\frac{eB}{m}(n+1/2))^2+
v^2p_z^2}$, and the extremal cross section occurs at $p_z=0$.  The appearance 
of the $1/2$ signals the absence of a finite Berry's phase and the expected 
trivial nature for oscillations. 

Considering now the opposite limiting case namely $\vartheta=\pi/2$  that is 
the magnetic field lying in the $x$ direction. Although the resulting Hamiltonian
cannot be diagonalized analytically, its spectral properties can be elucidated by
a semiclassical approach similarly to Ref. \onlinecite{montambaux}.
\begin{figure}
\includegraphics[width=0.9\columnwidth]{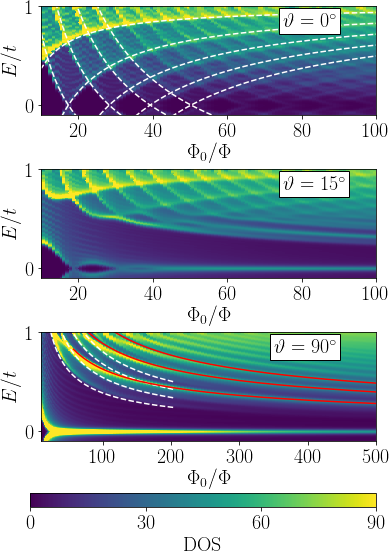}
\caption{Density of states for various magnetic filed orientations. Dashed white
lines for  $\vartheta=0$ stem from the solution of the Peierls substituted 
Eq. \eqref{hamnodalloop} with $\vartheta=0$,  while the dashed white and solid 
red lines arise from Eq. \eqref{Eapprox} using $\Phi/\Phi_0\sim B$ and $\Phi_0$ 
is the flux quantum. Already for a small $\vartheta=15^{\circ}$, the  features
from the topological oscillations are visible.}
\label{KPM_TB}
\end{figure}
This yields the spectrum, $E_{n,p_x}$, in the trivial and topological regimes 
from
\begin{subequations}
\begin{gather}
\frac{3\pi}{2\sqrt{2}}\left(n+\frac{1}{2}\right)=\sqrt{\epsilon} 
\left(2\alpha E(x)  +(\epsilon-\alpha)K(x)  \right),\\
\frac{3\pi}{2}n=\sqrt{\alpha+\epsilon}\left( \alpha E\left ({x}^{-1}\right)+
(\epsilon-\alpha)K\left ({x}^{-1}\right)\right),
\end{gather}
\end{subequations}
where $K(x)$ and $E(x)$ are the complete elliptic integrals\cite{abramowitz} of 
the first and second kind, respectively,
$x=(\epsilon+\alpha)/2\epsilon$, 
$\alpha=\tilde\Delta~(2m/(veB)^2)^{1/3}$, 
$\epsilon=E_{n,p_x}~(2m/(veB)^2)^{1/3}$, 
$n$ non-negative integer integer. The upper/lower equation corresponds to the 
trivial/topological part of the spectrum, respectively, and their boundary is 
at $\alpha=\epsilon$. The extremal parts of the spectrum (in $p_x$) for the 
topological and trivial regions read as
\begin{gather}
E_{n}\approx \left\{
\begin{array}{cc}
\pm 2\sqrt{\sqrt{\Delta}veBn/\sqrt{2m}}, \textmd{ topological}\\
\pm\left(\frac{veB}{\sqrt m}\left(n+\frac 12\right)\right)^{2/3}, 
\textmd{trivial},
\end{array}\right.
\label{Eapprox}
\end{gather}
signalling a Berry phase of $\pi$ and 0, respectively.
These correspond to the Landau levels of two dimensional massless 
Dirac\cite{castro07} and semi Dirac\cite{montambaux2009} points, respectively.

All these features, including the Berry phases, are reproduced from a numerical 
solution of Eq. \eqref{hamnodalloop} in a magnetic field, shown in Fig. 
\ref{inplanespectrum}. 
Apart from very small energies or chemical potential, the Onsager quantization 
approximates perfectly the numerically obtained spectrum.

\paragraph{Density of states.} 
The DOS (density of states) is given by $g(E)=\sum_{n,p_y,k}\delta\left(
E-E_{n,k}\right)$ with $k$ the appropriate quantum numbers for a given magnetic 
field orientation. For magnetic field in the $x-y$ plane, the $p_x$ dependence 
of the spectrum reveals that the contribution of the topological regions to the 
DOS overwhelms the trivial contribution  in this model.
The topological part has a much wider $p_x$ region close to its extremal point, 
therefore the curvature of the dispersion curve is much smaller, than that of 
the trivial region, where the curvature changes fast. From the conventional 
theory of magnetic oscillations\cite{abrikosov}, the DOS contribution of a 
given region is \emph{inversely} proportional to the curvature close to an 
extremal point, thus the topological contribution dominates the low energy part 
of the DOS. Upon increasing the magnetic field strength the topological 
oscillations shift up in energy and give way to trivial ones.

As we demonstrate in Fig. \ref{inplaneosc}, topological and trivial contributions
to the DOS can be disentangled. By following the evolution of the peaks in the 
DOS at fixed energy as a function of the inverse magnetic field, we find periodic
structures, as expected. These allow for the extrapolation back to the $n=0$
magnetic quantum number, revealing the underlying topological structure. Indeed,
in accord with our previous arguments, we find both topological (with Berry phase
$\pm\pi$, $\gamma=0$) and trivial (with Berry phase 0, $\gamma=\pm 1/2$) magnetic
oscillations, superimposed in top of each other.

In order to underline the robustness of our arguments we performed tight binding
calculations based on the lattice Hamiltonian
$H_\mathrm{TB}=\left (\delta-2t\sum_i \cos(k_i)\right)\sigma_x-
2t\sin(k_z)\sigma_z$, 
hosting a nodal loop as in Eq. \eqref{hamnodalloop}.
In all calculations we set $\delta=5t$ and a finite cubic lattice with 150 unit 
cells in each directions, was taken. Calculations were performed by employing 
the kernel polynomial method\cite{KPM,pybinding} including the magnetic field 
through a Peierls flux $\Phi$. The obtained DOS for various magnetic field 
orientations is depicted in Fig. \ref{KPM_TB}. These obtained oscillation 
patterns are in agreement with our previous arguments~\footnote{Only open source numerical packages such as numpy, scipy, matplotlib, pybinding, ipyvolume, were used to obtain the calculated results. We present our numerical code used, in form of jupyter notebooks at https://github.com/oroszl/nodalloopsemimetal}.

\paragraph{Conclusions.} 
In summary we have shown that the quantum oscillations in nodal loop semimetals 
exhibit a peculiar behaviour. While the oscillations for magnetic field 
perpendicular to the plane of the torus are trivial (i.e. with Berry phase 0), 
a transition occurs upon tilting the magnetic field, where the quantum 
oscillations consist of coexisting topological and non-topological oscillations 
with Berry phase $\pi$ and 0, respectively. This follows from an analysis of 
extremal Fermi surfaces, following the conventional theory of magnetic 
oscillations, supplemented with a topological inspection of the resulting 
cross sections. 
We emphasize that one can easily mix different (i.e. topological and trivial) 
magnetic peak sequences from the same band by looking at magnetic oscillation 
patterns in e.g. the DOS or magnetoresistance, and identify unphysical Berry 
phase contributions.
Therefore, it is of great importance to develop an analytical understanding,
such as our work, for analysing existing and future experiments on nodal line 
and other semimetals.

\begin{acknowledgments}
This research is supported by the National Research, Development and Innovation 
Office - NKFIH within the Quantum Technology National Excellence Program 
(Project No. 2017-1.2.1-NKP-2017-00001),  K105149, K108676, SNN118028, K119442, 
K115608,  K115575 and FK 124723 and by Romanian UEFISCDI, project number 
PN-III-P4-ID-PCE-2016-0032. A.C. acknowledges financial support through the 
MINECO/AEI/FEDER, UE Grant No. FIS2015-73454-JIN. and the Comunidad de Madrid 
MAD2D-CM Program (S2013/MIT-3007) L.O. acknowledges the Bolyai program of the 
Hungarian Academy of Sciences. Calculations were performed on the NIIF cluster.
\end{acknowledgments}

\emph{Note added:} Recently we became aware of a related 
works\cite{lim2,li2018}. Overlapping results are in agreement.

\bibliographystyle{apsrev}


\end{document}